# Title

Research Data Infrastructure for High-Throughput Experimental Materials Science

## Authors and Affiliations

Kevin R. Talley[1], Robert White[1], Nick Wunder[1], Matthew Eash[1], Marcus Schwarting[1,‡], Dave Evenson[1], John Perkins[1,#], William Tumas[1], Kristin Munch[1], Caleb Phillips[1], Andriy Zakutayev[1,*]

[1] Materials, Chemical and Computational Science Directorate; National Renewable Energy Laboratory, Golden, Colorado, 80401; USA
‡ Present address: Data Science and Learning Division; Argonne National Laboratory, Lemont, Illinois, 60439; USA
# Present address: Applied Chemicals and Materials Division; National Institute of Standards and Technology, Boulder, Colorado, 80305; USA
* Correspondence: Andriy.Zakutayev@nrel.gov

## Summary

The High-Throughput Experimental Materials Database (HTEM-DB) is the endpoint repository for inorganic thin-film materials data collected during combinatorial experiments at the National Renewable Energy Laboratory (NREL). This unique data asset is enabled by the Research Data Infrastructure (RDI) – a set of custom data tools that collect, process, and store experimental data and metadata. Here, we describe the experimental data-tool workflow from the RDI to the HTEM-DB to illustrate the strategies and best practices currently used for materials data at NREL. Integration of these data tools with the experimental processes establishes a data communication pipeline between experimental and data science communities. In doing so, this work motivates the creation of similar data workflows at other institutions to aggregate valuable data and increase its usefulness for future data studies. These types of investments can greatly accelerate the pace of learning and discovery in the materials science field, by making data accessible to new and rapidly evolving data methods.

## Keywords

data, metadata, materials, experimental, high-throughput, workflow

# Introduction

The High-Throughput Experimental Materials database (HTEM-DB)[Zakutayev-2017, Zakutayev-2018] is unique among commonly used materials databases[Stevanovic-2012, Jain-2013, Haastrup-2008, Curtarolo-2012, Saal-2013] because it contains experimental data rather than computational predictions. The broad and unfiltered dataset housed in the HTEM-DB is imported from ongoing experiments. Some databases contain useful experimental observations,[Mariette-2004, Xu-2011] but these are typically limited to only a specific collection of results (e.g. crystal structures) from published literature. Here, the complete experimental dataset is made available, including material synthesis conditions, chemical composition, structure, and properties. Databases similar to the HTEM-DB exist for a few scientific fields[Morrell-2017] and the need for such a resource has been recognized in others[Wenzel-2016]. While the workflow discussed here is based in custom data-tools, examples can be found using off-the-shelf data tools as well[Banko-2020].

HTEM-DB is enabled by a custom Research Data Infrastructure (RDI) - a modern data tool similar to a custom Laboratory Information Management System (LIMS). The RDI integrated into the experimental laboratory workflow that catalogues experimental data from inorganic thin-film materials experiments at the National Renewable Energy Laboratory (NREL). These long-term, ongoing experiments cover broad chemical ranges with high chemical resolution. This high-throughput experimental (HTE) process that feeds the HTEM-DB has been operating for more than a decade. It is fueled by materials studies which are motivated by factors unrelated to the HTEM-DB, yet provide a steady workflow for data mining. Collecting the results of experimental material synthesis and characterization creates a rich data source for machine learning studies.

Here, we focus on the custom RDI that has enabled the HTEM-DB at NREL. This article describes the structural pillars of this RDI, such as raw data collection, metadata collection, data extraction procedures, and data access, to document and discuss best practices for future data infrastructure projects of similar scope. After documenting the structural pillars, we discuss the impact of this workflow and the lessons learned during its implementation. While the example contained here employs high-throughput materials science studies, the overarching view is more broadly relevant to any experimental materials science laboratory working to bolster their data-related efforts. As such, this article aims to describe critical infrastructure developments that are relevant to experimental and data communities in materials science.

# Methods

In pursuit of a data infrastructure that augments and collects an experimental data stream for subsequent use by advanced algorithms, the first step is to analyze the needs of a specific research community with respect to its typical research cycle, as shown in the example of materials science in Figure 1. This analysis identifies both overlapping and unique data infrastructure requirements for each community, such as for the high-throughput experimental (HTE) and corresponding data communities discussed here. The HTE materials community begins a study by forming a hypothesis and testing it by experimentation. The data are processed and analyzed, and the results are reported by publication. The materials data community begins a study by identifying a set of relevant data. The dataset is then collected and sorted so that it can be filtered and analyzed for meaningful relationships, which are results reported by publication.

Each of these two cycles (experimental and data) has its own requirements, but the cycles can be integrated into a single workflow if the data needs are generalized as inflow, infrastructure, usability, or outflow. The infrastructure needs of each community are unique, where the experimental cycle requires tools for collecting, sorting, and storing newly generated data, whereas the data cycle needs easy access to the stored data. Differences in usability exist as well, where the experiment cycle needs tools to analyze and learn from the data, while the data community needs a large, diverse, high-quality data set. However, there is a strong overlap in the areas of inflow and outflow: access to previously obtained data and a controlled repository for new data are required for both cycles. These overlapping requirements for inflow and outflow motivated the creation of the. RDI, which collects, processes, and stores experimental data and metadata, and the HTEM-DB, which provides both a repository for experimental data and a data source for future data studies. The value of this workflow is increased when numerous, diverse studies populate the HTEM-DB with rich materials data, and when the resulting dataset is easily accessed and navigated for analysis by other materials researchers.

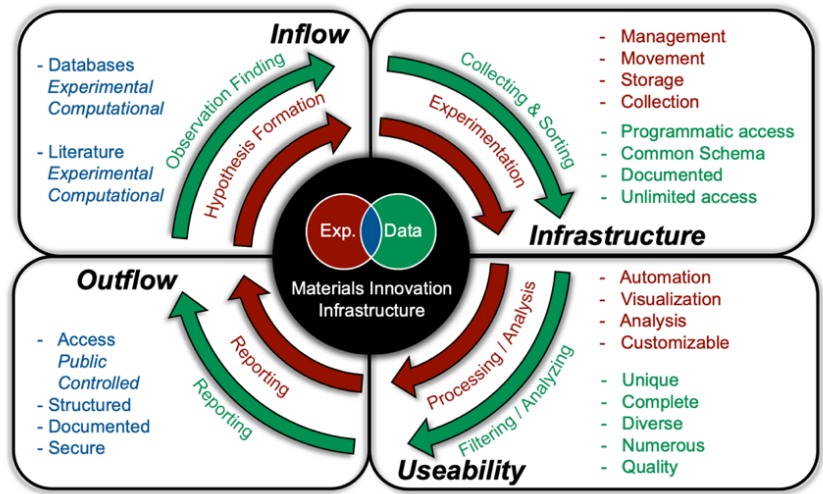

Figure 1. Materials data management needs for materials innovation, including experimental (red) and data (green) infrastructure. The experimental and data research cycles have (right) unique infrastructure and usability requirements, but (left) overlapping data inflow and outflow needs. These research cycles are combined within a common RDI and HTEM-DB, in which the existing experimental data stream is leveraged to develop new materials science insights through machine learning.

NREL possesses a wealth of high-throughput experimental capabilities and expertise for thin-film materials research. The experimental workflow (Figure 2) that is heavily utilized by numerous researchers involves depositing and characterizing thin films on 2" x 2" square substrates with a 4 x 11 sample mapping grid, which are common across multiple thin-film deposition chambers and spatially resolved characterization instruments. This experimental workflow at NREL has been benchmarked against other labs in Hattrick-Simpers *et al.*[Hattrick-Simpers-2019]. Other publications demonstrate the range of materials chemistries (e.g. oxides[Bikowski-2016], nitrides[Bauers-2020], chalcogenides[Siol-2018], Li-containing materials[Yu-2018], intermetallics[Zakutayev-2013]) and properties (e.g. optoelectronic[Welch-2017], electronic[Roberts-2020], piezoelectric[Talley-2018], electrochemical[Peng-2015]) to which these HTE methods have been applied. Each investigation generates large, comprehensive datasets that are harvested and delivered to the HTEM-DB by a custom RDI described in this paper. This present version of the RDI was first envisioned by Nelson *et al.*[Nelson-2003] almost two decades ago, then first described by White and Munch in 2014[White-2014], and briefly mentioned in Phillips *et al.*[Zakutayev-2018] in 2018.

The standard methods for producing, processing, measuring, and storing materials sample libraries (Figure 2, a-e) efficiently produce large amounts of data. In the past, these files were collected by individual researchers. Data management, processing, and storage responsibilities rested on these individuals. To address this burden, we created COMBIgor_Talley-2019—an open-source data-analysis package for high-throughput materials data loading, management, and visualization in combinatorial materials science. This data tool reduces the burdensome task of processing large data sets that result from high-throughput studies. The flexible platform enables experimental materials data to be efficiently handled, analyzed, visualized, and formatted for presentations, publications, and reports (Figure 2, f-h).

Another important component of the RDI is the NREL Research Data Network Warehouse, or Data Warehouse (DW) (Figure 2, center). The DW was first established at NREL in 2010, to manage data collected from laboratory computers that control experimental instruments. The DW collects data from these tools and makes the files accessible to researchers and to other data tools via the Research Data Network (RDN). For example, the HTEM-DB is populated with measurement data contained in specific high-throughput measurement folders in the DW that are identified by standardized file-naming conventions. Critical metadata from synthesis, processing, and measurement steps is also collected and added to the DW, providing experimental context for the measurement results. The data from these files is extracted, transformed, and loaded into the HTEM-DB, which stores processed data for analysis, publication, and data science purposes (Figure 2, right). The integration of this data workflow was made possible by the RDI that is detailed in the following sections.

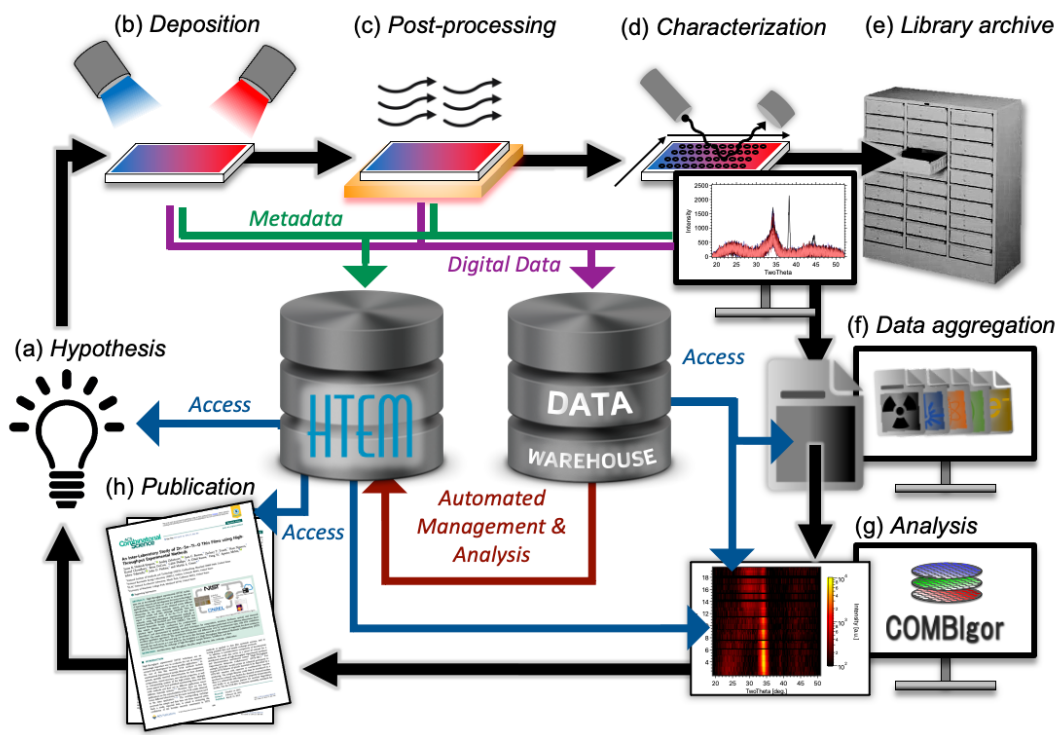

Figure 2. Experimental cycle for high-throughput materials discovery illustrating integration of the supporting data tools. Starting with (a) experiment design, material samples are (b) produced, (c) treated, and (d) measured and (e) stored in archives. The measurement data is (f) collected for (g) analysis and presented in (h) a publication, where it informs subsequent experiments. At each step, data tools were developed and implemented for the collection of (green) metadata and (purple) measurement data, (red) automated file management, and (blue) access. Here shown with data from Hattrick-Simpers *et al.*<sub>Hattrick-Simpers-2019</sub>.

## Results

The individual components of the RDI that facilitate the data workflow presented in Figure 2 form a set of interconnected, custom data tools. This includes tools for data collection (Data Harvesters and Laboratory Metadata Collector), data processing (Extract-Transform-Load), and storage and access (Data Warehouse and HTEM-DB). Brief descriptions of each of these data tools are provided below, and additional details can be found in the Supplementary Information.

*1. Digital data collection into the Data Warehouse*

Digital data is the primary source of materials data (Figure 3-a) within this data workflow(Figure 2). The RDI software harvests and stores all the digital files that are generated during materials growth and characterization processes. To keep the sensitive research instrumentation segregated from the normal NREL network activity, the computers are connected to the data harvester and archives via a firewall-isolated, specialized, sub network. The RDI system monitors activity on the instrument computers and identifies target files as they are created or updated. All relevant files on the instrument computers are copied into the DW archives and processed into the database as necessary.<sub>White-2014</sub> The DW contains data files that extend well beyond the scope of high-throughput materials experimentation, harvesting files from more than 70 instruments across labs and buildings on the NREL campus. The DW archives currently houses nearly 4 million files.

The DW consists of an object-oriented relational database and a Qumulo file system archive housed in the NREL Computational Sciences Data Center. The RDI codebase is built as a series of modules and libraries in C++ and bash scripts, but is architected to allow for easy replacement.<sub>White-2014</sub> The RDI architecture requires no software be installed on instrument computers. An additional advantage of this architecture is that it has no specific requirements for the individual instrument computers or file types; accommodating a wide range of computer operating systems and ages, typical of an experimental laboratory setting. The DW facilitates easy access to the resulting data files through a custom-built web application hosted on the internal NREL network, where the aggregated data files are easily accessed by researchers. Thus, the DW serves as both the central repository and the access gateway to experimental data generated at NREL.

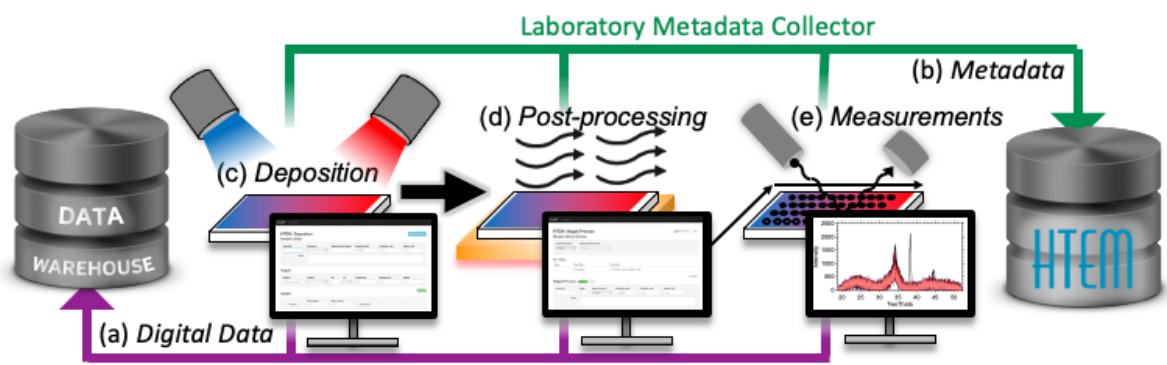

Figure 3. Data collection scheme. (a) Digital data, or raw data files, are collected by the RDI harvesters into the Data Warehouse while (b) contextualizing metadata is collected by the Laboratory Metadata Collector. Digital Data and Metadata are collected during three steps of the experimental process: (c) sample library growth, (d) sample post-processing, and (e) materials measurement. This data is combined and entered into the HTEM-DB as a complete sample record. Detailed images of the custom web forms are contained in the supporting information.

2.  *Metadata collection by the Laboratory Metadata Collector* (Figure 3,b)
Metadata is one of the most critical types of data in this workflow (Figure 2) because it gives context to the data files collected in the DW. This data stream, although of high importance, is difficult to capture because it requires interaction and input from humans (experimentalists). To simplify and streamline metadata collection, the Laboratory Metadata Collector (LMC) was developed and implemented. The LMC includes custom webforms in which users enter and submit sample record information upon completion of an experiment. Each record in the LMC is a detailed, digital lab notebook entry, giving researchers and algorithms access to the experimental variables of interest and making them easily comparable to measurement data stored alongside metadata in the HTEM-DB.

The LMC is comprised of a front-end, single-page web application written in JavaScript and a back-end application programming Interface (API) written in Node.js. It is hosted on a VM located in the same NREL data center as the DW and writes to the existing PostgreSQL database that houses DW and HTEM data. The bulk of the logic is in the front-end web application, which runs in any modern web browser. It can be accessed from the NREL network by researchers' laptops and by in-lab computers. The application provides model logic for building webforms and conducts form validation. The dynamic view of the webform is adjusted in response to user-entered values and the user's display settings. It presents a preview of the entry upon submission for the user to verify and provides the option for the user to either save the entry or return to editing. The front-end also provides a searchable interface for finding previous entries. Entry submission and search features are enabled through a back-end API that interacts with the DW. The API simply adds the submitted JSON data to the DW and queries and retrieves past entries.

The LMC webforms collect information from the experimentalist during deposition, post-processing, and characterization of sample libraries. This information is agglomerated in the HTEM-DB for each sample included on the form, immensely enriching the value of the associated measurement data. Providing additional benefit to the user, COMBIgor plugins port this LMC-generated information directly into the user's Igor Pro experiment via direct JSON file download/import or through the HTEM API. Easy access to complete experimental records in COMBIgor motivates researchers to participate in metadata reporting through the LMC. As such, the LMC supports the data needs of both the experimental and data communities. For the experimentalist, it provides a more efficient, accurate, and accessible record of experimental variables for experimental analysis, compared to typical handwritten lab notebooks. For the data community, the LMC offers reliable and complete sample records for individual material samples, which increases the value of any associated data and the overall usability of the entire HTEM-DB in larger data studies.

3. *Data extraction, transformation, and loading* (Figure 4, d-f)

A significant number of resources has been used to develop the custom extract, transform, load (ETL) scripts that port raw data from files in the DW into the PostgreSQL HTEM-DB. These scripts, written primarily in Python, run on the NREL high-performance computing system, and are automatically executed daily.

First, (1) the folders of high-throughput data are harvested from the DW and filed into the HTEM-DB repository, based on file name conventions, using a Python script. The information is extracted from the raw files in these folders and placed into tables that correspond to the standard library mapping grid from HTE studies at NREL. Next, (2) the data extracted from the files is transformed from original (sometimes proprietary) to final format. The extraction method varies depending on the specific file type and, in some cases, utilizes additional Python libraries or programming languages (e.g., ruby, C, R). Finally, (3) individual pieces of new data for a given sample are identified and loaded to the PostgreSQL database. In this way, these data are correlated to other descriptive data pertaining to the same sample.

Collectively, the ETL process of (1) extracting, (2) transforming, and (3) loading the data from the DW produces a continuous flow of new information to the HTEM-DB and supports both experimental and data communities. As part of the ETL workflow, certain data types may be combined to generate additional, useful data types. For example, sheet resistance and thickness may be combined to calculate resistivity, or optical transmission, reflection, and thickness may be combined to calculate the absorption spectrum and to determine the optical bandgap[Schwarting-2017]. Thus, this ETL infrastructure funnels and correlates datasets from raw files in the DW into structured database entries in the HTEM-DB that are easily accessed by experimental and data-focused researchers.

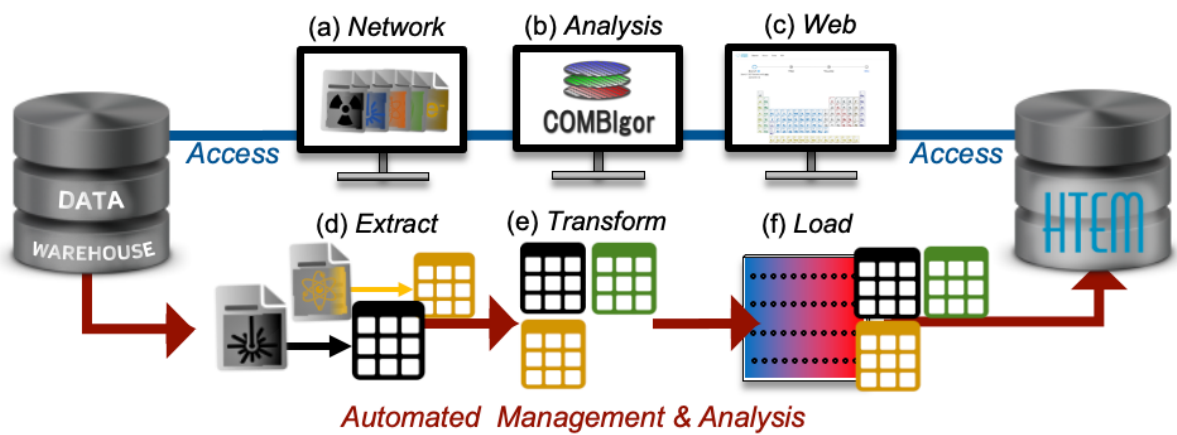

**Figure 4**. Data management scheme. For experimental researchers, the raw files can be obtained from the DW via (a) network access and can be easily loaded for (b) flexible analysis of the data in COMBIgor. Curated sample records can be searched, filtered, visualized, and downloaded through (c) HTEM-DB web access. A custom plugin was developed to load sample records through the HTEM automated programmatic interface into a user analysis environment in COMBIgor. To populate the HTEM-DB with sample records, raw measurement files from the DW are processed through a set of custom ETL scripts. The data is (d) extracted from files in the DW, (e) transformed into useable data, and (f) loaded to the corresponding sample library entries. In certain cases, the loaded raw data are also post-processed into derived data before being added to the HTEM-DB.

4.  High-Throughput Experimental Materials Database (HTEM-DB) (Figure 4)

The HTEM-DB, including its structure, content, and applications, is detailed in Zakutayev et al.$_{\text{Zakutayev-2018}}$. Here, we present the HTEM-DB in a level of detail similar to the other workflow components in Figure 4 for completeness. The HTEM is a PostgreSQL database that is the reservoir for all the incoming data from experimental workflows. It is housed in the same NREL data center as the DW, LMC, and ETL scripts. This is the most visible part of the RDI because the HTEM-DB functions as an access point for both experimental and data communities. There are two main points of access to the rich information that the HTEM-DB$_{\text{Zakutayev-2017}}$ contains. Experimentalist typically use the (1) web-interface (https://htem.nrel.gov) which provides the user with quick and effective sample search for retrieval of sample and library information of all data types and dimensionalities, which can then be viewed using the built-in data visualization features. Data scientists typically use the (2) API (https://htem-api.nrel.gov/api), which enables machine learning by providing algorithms with programmatic access to the entire HTEM. The API is also used to load experimental data into COMBIgor for further visualization and manipulation. Thus, the HTEM-DB supports both communities by providing tailored data inflow and outflow access and by providing features that enhance the usability of the data that it contains.

Access to the data that flows through RDI provides researchers with an opportunity to interact with it and learn from it. For researchers conducting experimental studies, and generating the continuous flow of incoming materials data, the RDI described here provides improved efficiency and increased accuracy of experimental research data handling. To this material science community, the automated data workflow provides easy access to aggregated data for samples of interest in ongoing research projects. This can be achieved through downloading files directly from the DW, loading library information into COMBIgor directly from the HTEM-DB or, viewing the data directly using the HTEM web interface. The data analysis tools built into the HTEM web interface for finding, filtering, and visualizing sample records give unique opportunities to explore the dataset in support of a variety of research goals. For the data scientist, who is analyzing the larger HTEM dataset, the RDI provides a physically based dataset to investigate by machine learning methods. To the data science community, the RDI provides access to curated, structured, and complete data records for a wide range of materials. This is achieved through the HTEM API for the publicly available data. Access to the entire contents of the HTEM is controlled in the interest of all the stakeholders, including the public or private funding sources who support the work and demand exclusive access to the resulting data.

The HTEM-DB is large, mostly complete, and diverse (Figure 5, a-c). Due to a rich history of HTE materials studies conducted at NREL, the database has been populated with more than 300,000 unique samples from more than 7,000 sample libraries. Of the unique samples within the HTEM-DB, more than half have correlated characterization data. These samples cover a wide range of compositions, with more than 33 elements quantified in samples with composition measurements. These characteristics are desirable for materials data studies and are important factors if such studies are to be realistic and useful. The HTEM-DB will continue to grow due to ongoing experiments and continuing efforts to build additional collection pathways. With the continued growth of the database, there will be an ever-increasing potential for useful knowledge extraction.

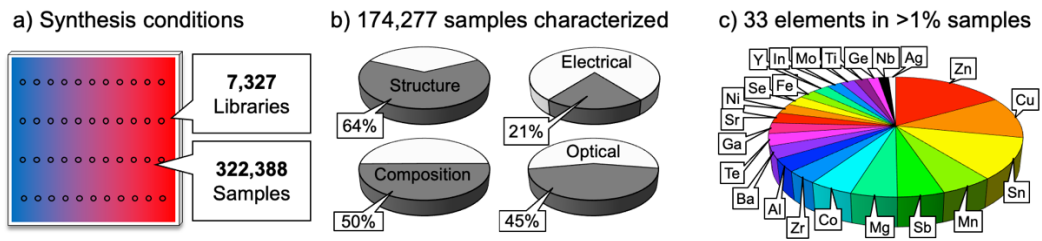

Figure 5. HTEM database statistics. As of September 2020 the database is (a) large, with over 300,000 unique samples on more than 7,000 libraries, (b) complete with either structure, composition and/or property data for more than 170,000 samples, and (c) chemically diverse, shown as percent of samples containing a given element.

## Discussion

The effort to develop the RDI at NREL has resulted in a more valuable product than initially envisioned. By establishing various RDI tools, integrating them with each other, and providing them to researchers, a complete data workflow has been constructed that supports and complements the existing experimental research cycle while curating valuable data for future use in data studies. These RDI tools are tailored to the specific experimental setting at NREL, but are broadly designed to meet the needs outlined in Figure 1-a. Thus, this RDI example should be applicable to other materials science laboratory settings with a somewhat standardized research process. The data tools that form the RDI— specifically the harvesters, DW, the ETL process, the LMC, and the HTEM-DB — are all critical to the success of the RDI. The hope is that the designs and features of the data tools described here will serve as examples of best practices for other institutions that require similar RDI for their own data workflows.

The modern materials science laboratory is a difficult and complex terrain to capture by a cohesive set of data support tools. Experimental equipment containing all types of software, hardware, age, and access will be encountered. This requires skilled and dedicated resources in material science, data science, and software engineering. A RDI built to collect, maintain, and access the data should be simple, to reduce the maintenance cost and simplify efforts to scale. Ideally, these types of systems should be engineered from the top-down rather than the bottom-up, requiring significant investments in infrastructure from the beginning. Building a complete data workflow in this terrain requires a substantial financial investment in hardware, network installation, software development, and maintenance. However, the RDI at NREL has been constructed from the bottom up with multiple contributions from many people over the time span of more than a decade. For example, the initial version of RDI has been prototyped as a part of a laboratory design and construction project by DOE, and more recently supported by NREL internal research data infrastructure initiative. The resulting RDI products are functional, though not ideal, and can be used to make a blueprint for better RDI planning and construction in the future. In a similar way, designing and constructing the HTEM-DB has not been externally funded as a directed effort, but rather as a collective consequence of addressing individual data challenges in the beginning of an internally-supported cross-collaborative relationship between materials science and data science. The resulting materials-data relationship has shown promise as a valuable contributor to materials science, and as such, encourages investments in this field to design and build similar data-workflows and databases at other institutions.

The most successful data tools to prompt user engagement at NREL were built through tight collaborative design efforts between material science users and data scientists and software engineers. The RDI constructed by software engineers and presented here is useful to the experimentalist and, as a result, has been widely adopted by the materials researchers at NREL. As more materials researchers use the system, the HTEM-DB continues to collect, preserve, and provide materials data. In turn, the large, curated set of samples synthesized and characterized by materials scientists is primed and ready for exploration by machine learning algorithms by the data scientists. This RDI and HTEM-DB are just one example of how individual materials science and data science effort at NREL provide much greater value when combined to work together. Similar systems brought online at other institutions, producing a broader set of materials data to explore, would further increase the potential of this materials-data relationship to advance science.

## Acknowledgments

This work was authored at the National Renewable Energy Laboratory, operated by Alliance for Sustainable Energy, LLC, for the U.S. Department of Energy (DOE) under Contract No. DE-AC36-08GO28308. Financial support for the RDI operation and improvements, including LMC, is provided by NREL indirect funding. Funding HTEM DB development was provided by several NREL's Laboratory Directed Research and Development (LDRD). HTEM data curation efforts were funded by U.S. Department of Energy (US DOE), Office of Science. Data Warehouse prototyping was supported by the U.S. Department of Energy (US DOE), Energy Efficiency and Renewable Energy (EERE). A portion of the research was performed using computational resources sponsored by the Department of Energy's Office of Energy Efficiency and Renewable Energy and located at the National Renewable Energy Laboratory.

## Author Contributions ([CRediT](#))

Conceptualization, A. Z. and C. P.; Software, R. W., N. W., C. P., D. E., M. E.; Resources, B. N., C. P., K. M., J. P., A. Z.; Data Curation, R. W., M.S., C. P., A. Z.; Writing – Original Draft, K. R. T.; Writing – Review and Editing, K. R. T., C. P., and A. Z.; Visualization, K. R. T.; Supervision, K. M., C. P., B. N., A. Z.; Project Administration, K. M, B. N., C. P., J. P., A. Z. ; Funding Acquisition, B.N., J. P., K. M., C. P., A. Z.

## Declaration of Interests

The authors declare no competing interests.